\documentclass[12pt]{article}

\usepackage[english]{babel}
\usepackage[latin1]{inputenc}

\usepackage{latexsym,amsfonts,amsmath,amsthm,amssymb,graphicx}

\title{Distributivity breaking and macroscopic quantum games}

\author{A. A. Grib, G. N. Parfionov and K. A. Starkov\\
Alexandre Friedmann Laboratory of Theoretical Physics, \\
St.Petersburg University of Economics and Finances, Russia\\
A. Yu. Khrennikov\\
International Center for Mathematical Modeling\\
in Physics, Engineering and Cognitive science\\
MSI, V\"axj\"o University, S-35195, Sweden}

\begin{document}

\maketitle

\begin{abstract}
Examples of games between two partners with mixed strategies,
calculated by the use of the probability amplitude as some vector
in Hilbert space are given. The games are macroscopic ,no
microscopic quantum agent is supposed. The reason for the use of
the quantum formalism is in breaking of the distributivity
property for the lattice of yes-no questions arising due to the
special rules of games. The rules of the games suppose two parts:
the preparation and measurement. In the first part due to use of
the quantum logical orthocomplemented non-distributive lattice the
partners freely choose the wave functions as descriptions of their
strategies. The second part consists of classical games described
by Boolean sublattices of the initial non-Boolean lattice with
same strategies which were chosen in the first part. Examples of
games for spin one half are given. New Nash equilibria are
found for some cases. Heisenberg uncertainty relations without the
Planck constant are written for the "spin one half game".\end{abstract}

\section{Introduction}\label{intro}

The aim of this paper is to give examples of macroscopic games
between two agents, call them Alice and Bob, described by the
formalism of  quantum mechanics for some simple spin one half
cases.

The games are macroscopic which means that differently from widely
discussed now in literature [1] of the so called "quantum
games" no  microscopic agent like molecule, atom or photon
necessarily described by quantum physics is supposed to be
present.

The idea is to look for macroscopic situations in which random
behavior cannot be described by the standard Kolmogorovian
probability measure but by the probability amplitude represented
by some vector in Hilbert space(see wide discussion of the problem
in papers of A. Yu. Khrennikov and his colleagues on V\"axj\"o
conferences [2]--[8].

Games look natural candidates for such situations. Our search for
such situations is motivated by the discovery of Birkhoff and von
Neumann [9] that the reason for the different description of
chance in quantum physics is in the non-distributivity property of
the lattice of properties of the microscopic system. As it is well
known these lattices for simple spin one half and spin one systems
are quantum logical orthocomplemented lattices.

For breaking the distributivity rule it is enough to look for
situations in which disjunction for exclusive cases is defined not
uniquely. The first examples of such situations for macroscopic
automata were given in [10], [11].
Later in the papers [12],  [13] first examples of
such games imitating quantum spin one half and spin one particles
with two non commuting observables measured were given.

The mathematical rule of formulating the quantum game is simple
and natural. One takes some classical game with the given payoff
matrix and mixed strategies (in paper [12] it was called "the
foolish Alice game") and in the expression for the average profit
depending on the payoff matrix and probabilities of the chosen
strategy puts instead of the characteristic functions of Alice
questions and Bob's answers in payoff function "operators",
represented as projectors in some Hilbert space. The average
profit is calculated as the mathematical expectation value of the
"payoff "operator for the tensor product of wave functions
representing now the strategies of Alice and Bob.

Nash equilibria are found concerning different choice of wave
functions by Alice and Bob. Different representations of the
lattice by projectors depending on choice of observables by Alice
and Bob, parameterized for the spin case by some angle between
projections of spin observables are considered. New Nash
equilibria are found.

However the most interesting problem is interpretation of such
games for classical situations. In some sense this is similar to
the problem of "hidden variables "in quantum physics, where one
tries to find some classical system, which "effectively" is
described by the formalism of quantum mechanics [14].

In this paper to clarify the rules given in [12], [13] and to
come closer to the "quantum casino" realization of the games one
adds a new rule added to those given in the cited papers.

One must divide the game on two parts:

\medskip

1. The preparation part, the rules of which
    are similar to those given in [12], [13].
\medskip

2. Measurement of two or more non-commuting operators
    of the described system. This second part consists of two or more classical games,
    the strategies in which must be those chosen by the partners in the first part.
    This "must" means some following of the "tradition" chosen in the first part.

\medskip

In standard quantum mechanics as it is known the frequencies of
the results for measurements of different non-commuting
observables with the definite prepared wave function are
predetermined and cannot be arbitrary.

From the point of the axiomatic quantum theory as theory of
quantum logical lattices Part 2 of our games corresponds to taking
distributive sublattices of the initial non-distributive lattice
with values of frequencies (or classical probabilities) prescribed
by the quantum probabilistic measure (the wave function).

Preparation of wave functions of Alice and Bob means defining
frequencies of definite exclusive positions of Bob's ball and
Alice's questions. These frequencies however differently from
classical games have less freedom in their definition. For example
in our first game imitating the spin one half system with two
non-commuting observables Alice and Bob due to constraint for
frequencies by the wave function have free choice only defining
freely one angle.

Besides the examples considered in our previous papers an example
of spin one half system with three non-commuting observables of
spin projections is concerned. For this case one can look for the
imitation of Heisenberg uncertainty relations for spin projections
in the macroscopic quantum game.

Let us describe these rules more explicitly

\bigskip

1. {\it The preparation stage.}

\medskip

Alice and Bob have two quadrangles, one for Alice another for Bob.
As in [12], [13]. Bob puts the ball in some vertex of his
quadrangles. Alice exactly guesses to which vertex Bob put his
ball. She does this by asking questions: is the ball in the vertex
"a"? However Bob gives the answer "yes" not only in case the ball
is in "a" but also in cases if the ball is in vertices connected
by one arch with "a". It is only if the ball is in the opposite
vertex that he cannot move it and definitely answers "no". This
means that only "negative" answers of Bob are non-ambiguous for
Alice. Alice on stage 1 fixes the number of non-ambiguous answers
of Bob and calculates some frequencies for opposite vertices:
 $$\omega_{1,B}=\frac{N_1}{N_1+N_3},\quad \omega_{3,B}=\frac{N_3}{N_1+N_3},$$
 $$\omega_{2,B}=\frac{N_2}{N_2+N_4}, \quad \omega_{4,B}=\frac{N_4}{N_2+N_4}$$

Now to do the game symmetric the same in supposed for Alice. Alice
puts her ball to some vertex of the quadrangle and Bob must
exactly guess the vertex. Only negative answers are non-ambiguous
for Bob.
\begin{figure}[ht]
\begin{picture}(200,90)
\put(5,68){\bf 2} \put(70,68){\bf 3}

\put(5,5){\bf 1} \put(70,5){\bf 4}

\put(30,30){\bf Alice}

\put(145,68){\bf 2} \put(210,68){\bf 3}

\put(145,5){\bf 1} \put(210,5){\bf 4}

\put(170,30){\bf Bob}

\linethickness{2pt}

\put(0,0){\line(1,0){80}}

\put(140,0){\line(1,0){80}}

\put(0,80){\line(1,0){80}}

\put(140,80){\line(1,0){80}}

\put(0,0){\line(0,1){80}}

\put(80,0){\line(0,1){80}}

\put(140,0){\line(0,1){80}}

\put(220,0){\line(0,1){80}}

\end{picture}
\vspace{-15pt}\caption{\em } \label{fig1}
\end{figure}
To any graph of Fig. \ref{fig1} corresponds non-distributive
lattice Fig. \ref{fig2}.
\begin{figure}[ht]
\begin{picture}(120,120)
\put(45,94){\circle*{4}} \put(43,105){I} \put(0,50){\circle*{3}}
\put(-10,55){1} \put(30,50){\circle*{3}} \put(25,55){2}
\put(60,50){\circle*{3}} \put(65,55){3} \put(90,50){\circle*{3}}
\put(95,55){4} \put(45,6){\circle*{4}} \put(42,-10){O}
\put(0,50){\line(1,1){45}} \put(0,50){\line(1,-1){45}}
\put(90,50){\line(-1,1){45}} \put(90,50){\line(-1,-1){45}}
\put(30,50){\line(1,3){15}} \put(30,50){\line(1,-3){15}}
\put(60,50){\line(-1,3){15}} \put(60,50){\line(-1,-3){15}}
\end{picture}
\vspace{-15pt}\caption{\em } \label{fig2}
\end{figure}

Definite frequencies $\omega_{a,B}$ mean that the wave function of
Bob's ball is given as   $\psi_B$ and the representation of the
lattice (Fig. \ref{fig2}) is defined, so that two pairs of
orthogonal projectors $\hat p_1$, $\hat p_3$, $\hat p_2$, $\hat
p_4$ are chosen. Then $\omega_{a,B}=\langle\psi_b|\hat
p_a|\psi_b\rangle$ as it must be for the quantum spin $\frac12$
system with two observables -- spin projections $\hat S_z$   and
$\hat S_\theta$ with $\theta$ some angle being measured. Definite
frequencies $\omega_{a,A}$ define the wave function of Alice's
ball and some observables $\hat S_z$, $\hat S_\theta$ for Alice.

\bigskip

{\it 2. Measurement stage.}

\medskip

Two classical games are considered.
Bob puts his ball only to vertices on one diagonal in the first
game, let it be 1, 3. Alice asks questions trying to guess the
position of Bob's ball. However, the frequencies of putting by Bob
his ball to some vertices "must" be $\omega_{1,B}$, $\omega_{3,B}$
defined on the first stage. The frequencies of Alice's questions
"must" be $\omega_{1,A}$, $\omega_{3,A}$. Money are paid to Alice
on this stage and their amount is fixed by the payoff matrix. In
the second game Bob put the ball to vertices 2, 4 with frequencies
$\omega_{2,B}$, $\omega_{4,B}$ and Alice asks questions with
frequencies $\omega_{2,A}$, $\omega_{4,A}$. The profits in two
games are added.

The result will be given by use of the expectation value of the
sum of projectors multiplied on the elements of the payoff matrix
for the tensor products of two wave functions, as it was written
in [12], [13]. The quantum game so formulated is the irrational
game which makes it's theory different from the usual game theory.
The payoff matrix is known to the players from the beginning, but
as we see on the "measurement" stage it doesn't motivate their
behavior. However it can motivate somehow their behavior on the
first stage when the wave functions and "observables" are defined.
Nash equlibria for fixed angles for observables can be understood
as some "patterns" in random choice of two players.

\section{Spin one half game with three observables}

Here we consider more complicated game imitating particle with
spin one half, for which three non-commuting observables $\hat
S_x$, $\hat S_y$, $\hat S_z$ are measured. This case is
interesting because here one can imitate Heisenberg uncertainty
relations for spin projections in case of our quantum game. For
simplicity we consider that same observables are measured by Alice
and Bob (no difference in angles between projections is supposed).
The Hasse diagram for this case is:

\begin{figure}[ht]
\begin{picture}(110,100)
\put(60,10){\circle*{3}} \put(60,-5){$O$}

\put(10,40){\circle*{3}} \put(2,35){$1$}

\put(30,40){\circle*{3}} \put(23,35){$2$}

\put(50,40){\circle*{3}} \put(43,35){$3$}

\put(70,40){\circle*{3}} \put(72,35){$4$}

\put(90,40){\circle*{3}} \put(92,35){$5$}

\put(110,40){\circle*{3}} \put(112,35){$6$}

\put(60,70){\circle*{3}} \put(60,83){$I$}

\put(60,10){\line(-5,3){50}} \put(60,70){\line(5,-3){50}}

\put(60,10){\line(-1,1){30}} \put(60,70){\line(1,-1){30}}

\put(60,10){\line(-1,3){10}} \put(60,70){\line(1,-3){10}}

\put(60,10){\line(1,3){10}} \put(60,70){\line(-1,-3){10}}

\put(60,10){\line(1,1){30}} \put(60,70){\line(-1,-1){30}}

\put(60,10){\line(5,3){50}} \put(60,70){\line(-5,-3){50}}

\end{picture}
\vspace{-15pt}\caption{\em } \label{fig3}
\end{figure}
\noindent Here orthogonal projectors are $1-4, 2-5, 3-6$ which
correspond for the spin $\frac 12$ case to $\hat S_x=\pm \frac
12$, $\hat S_y=\pm \frac 12$, $\hat S_z=\pm \frac 12$.

\noindent The rule is the same, Bob can move his ball on one step depending
on Alice question. For example he can move to 1 from 2, 6, 3, 5
but not from 4 etc. For Alice "no answer" on 1 means "Bob is at
4", "no answer on 2" means he is at 5 etc. Same is supposed to
Alice's ball and Bob's questions. Representation of atoms of the
Hasse diagram~(Fig.~\ref{fig3}) by projections is:
\[ A_1=\begin{pmatrix}
  1 & 0 \\
  0 & 0
\end{pmatrix}\quad
A_2=\frac 12 \begin{pmatrix}
  1 & 1 \\
  1 & 1
\end{pmatrix}\quad
A_3=\frac 12 \begin{pmatrix}
  1 & -i \\
  i &\ 1
\end{pmatrix}\quad
\]

\[
A_4=\begin{pmatrix}
  0 & 0 \\
  0 & 1
\end{pmatrix}\quad
A_5=\frac 12\begin{pmatrix}
  1 & -1 \\
  -1 & -1
\end{pmatrix}\quad
A_6=\frac 12\begin{pmatrix}
  1 & i \\
  -i & 1
\end{pmatrix}
\]

\newpage

The graph
of the game, showing vertices to which Bob and Alice put their
balls is the following.

\begin{figure}[ht]
\begin{picture}(120,120)

\put(0,50){\circle*{3}}\put(-10,55){1}
\put(0,50){\line(1,0){100}}\put(0,50){\line(0,-1){50}}\put(0,50){\line(3,-5){50}}
\put(0,50){\line(3,2){50}}

\put(50,-35){\circle*{3}}\put(50,-50){5}
\put(50,-35){\line(-3,2){50}}\put(50,-35){\line(3,2){50}}

\put(100,50){\circle*{3}} \put(105,55){3}
\put(100,50){\line(0,-1){50}}\put(100,50){\line(-3,-5){50}}\put(100,50){\line(-3,2){50}}

\put(100,0){\circle*{3}} \put(105,5){4}
\put(100,0){\line(-3,5){50}}

\put(50,84){\circle*{3}} \put(50,90){2}

\put(0,0){\circle*{3}}\put(-10,5){6}
\put(0,0){\line(1,0){100}}\put(0,0){\line(3,5){50}}

\end{picture}

\end{figure}

\bigskip

\bigskip

\bigskip

\medskip

\medskip

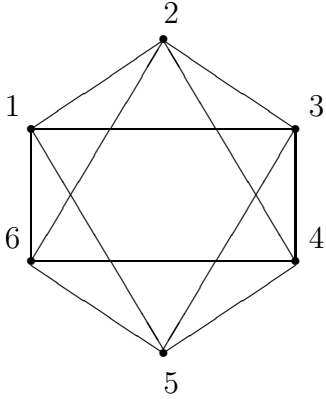
\centerline{\bf Figure 4.}

The payoff matrix is:

\bigskip

\bigskip

\begin{table}[h]
\caption{The Payoff-matrix of Alice}\label{h11}
\begin{tabular}{|c||c|c|c|c|c|c|}  \hline
  $A \backslash B$ & 1 & 2 & 3 & 4   & 5 & 6 \\ \hline\hline
  1 & 0 & 0 & 0 & $v_1$ & 0 & 0 \\ \hline
  2 & 0 & 0 & 0 & 0 & $v_2$ & 0 \\ \hline
  3 & 0 & 0 & 0 & 0 & 0 & $v_3$ \\ \hline
  4 & $v_4$ & 0 & 0 & 0 & 0 & 0 \\ \hline
  5 & 0 & $v_5$ & 0 & 0 & 0 & 0 \\ \hline
  6 & 0 & 0 & $v_6$ & 0 & 0 & 0 \\ \hline
\end{tabular}
\end{table}

\noindent Then the payoff operator of the quantum game is
$$\hat P=v_1 A_1\otimes B_4+v_2 A_2\otimes B_5+v_3
 A_3\otimes B_6+v_4 A_4\otimes B_1+v_5
A_5\otimes B_2+v_6 A_6\otimes B_3
$$
Here all $B_i$ for Bob has the same form as $A_i$. The strategies of Alice asking questions
and Bob putting the ball in their graphs (Fig.4) are
described as frequencies of choices in "preparation part" given by
wave functions, represented as vectors in complex Hilbert space: $
\varphi_A=(\cos\alpha, e^{i\theta}\sin\alpha)$,
$\psi_B=(\cos\beta, e^{i\omega}\sin\beta) $. So generally
differently from real two dimensional space in the previous
example [12], [13] one can take as in quantum spin one half
physics complex space. The average profit in subsequent three
"measurement" games is: $$
E_A=\langle\varphi_A|\;\otimes\langle\psi_B|\;\hat
P\;|\psi_B\rangle\otimes\;|\varphi_A\rangle.$$ It is calculated as
\[
E_A=v_1\cos^2\alpha\sin^2\beta+v_2 \frac{1+\cos\theta\sin
2\alpha}{2}  \cdot\frac{1-\cos\omega\sin 2\beta}{2}  +\]
\[+ v_3 \frac{1+\sin\theta\sin 2\alpha}{2}
\cdot\frac{1-\sin\omega\sin 2\beta}{2}
+v_4\sin^2\alpha\cos^2\beta+
\]
\[+ v_5 \frac{1-\cos\theta\sin 2\alpha}{2}
\cdot\frac{1+\cos\omega\sin 2\beta}{2} + v_6
\frac{1-\sin\theta\sin 2\alpha}{2} \cdot\frac{1+\sin\omega\sin
2\beta}{2}
\]
So Nash equilibria can be found by analyzing the function
$E_A(\alpha,\beta,\theta,\omega)$. The simplest case is when
$\theta=\omega=0$ and $\varphi_A$, $\psi_B$ are real. For this
case define  \[a=v_1,\; b=v_4,\; c=-\frac{v_2+v_3+v_5+v_6}{4},\;
d=\frac{v_2+v_3-v_5-v_6}{4}\] then $E_A=H(\alpha,\beta)$, where
\[H(\alpha,\beta)=a \cos^2 \alpha \sin^2 \beta+b \sin^2 \alpha
\cos^2 \beta+c(1-\sin 2\alpha \sin 2\beta)+ d(\sin 2\alpha-\sin
2\beta)\] To find Nash equilibria one must look for intersection
points of curves of reaction of Bob and Alice [12]. Three
cases were investigated by us.
\begin{itemize}
  \item a=7, b=1, c=-2, d=1,5. Nash equilibrium exists for
  $\alpha=\beta=\frac{\pi}{8}$.
  The value of the payoff at this point is equal to 2.
  \item a=1, b=1, c=-2, d=0. No Nash equilibrium exists for this
  case.
  \item a=1, b=10, c=-6, d=4. Nash equilibrium exists for $\alpha=87,9^0$,
  \mbox{$\beta=69,2^0$}, $E_A=4,6$.
\end{itemize}

\section{Heisenberg's uncertainty relations}
As we said before, the game consists of two parts:
\begin{enumerate}
\item "preparation"when the non-distributive quantum logical lattice was used, leading
        to  the choice of Alice and Bob of their wave functions.
\item measurement, described by three different games, using orthogonal
        vertices  of the graph~(fig.\ref{fig3}) and described by frequencies obtained from part 1.
\end{enumerate}
As it is well known from quantum mechanics there are Heisenberg's
uncertainty relations for spin projections, so that if
$$
[\hat S_x,\hat S_y]=i\hbar\hat S_z,
$$
then for dispersions one has
\begin{equation}
\label{H1}
D_\psi S_x \cdot D_\psi  S_y\geqslant \frac{\hbar^2}{4}(E_\psi
S_z)^2
\end{equation}
As it was shown in [10], [11] earlier
these relations for graph are equivalent to relation for
frequencies obtained from the wave function:
\begin{equation}
\label{H2}
p_1 p_4 p_2 p_5\geq\frac{1}{16}(p_3-p_6)^2
\end{equation}
 Here
$p_i$-frequencies for Alice. Same relation is valid for Bob. In
our case
\[
p_1=\cos^2\alpha,\quad p_2=\frac{1+\cos\theta\sin
2\alpha}{2},\quad p_3= \frac{1+\sin\theta\sin 2\alpha}{2},
\]
\begin{equation}
\label{H3}
p_4=\sin^2\alpha, \quad p_5=\frac{1-\cos\theta\sin 2\alpha}{2}
,\quad p_6=\frac{1-\sin\theta\sin 2\alpha}{2}
\end{equation}
Then (\ref{H2}) means $ \sin^2 2\alpha\leq 1$ which is always valid.

In our case of three classical games with probabilities prescribed
by (\ref{H3})one can consider measuring three random variables
 $A_1$, $A_2$, $A_3$ taking values $\pm 1$ and calculate dispersions and expectation values
$$
D(A_1)=\sin^2 2\alpha,\qquad D(A_2)=1-(\cos\theta\sin 2\alpha)^2,
\qquad E(A_3)=\sin\theta\sin 2\alpha
$$
So one obtains
\begin{equation}
\label{H4}
D(A_1) D(A_2)\geq (E(A_3))^2
\end{equation}
equivalent to (\ref{H2}). Here differently from(\ref{H1}) we put $\hbar=1$
and there is no $\frac 12$ as it was for spin variable. However if
one includes in the notion of observable the "payment" defined by
the payoff matrix, then for all equal $v$ in the payoff matrix one
can see that dimensional "price" can play the role of the
Planckean constant.

\section{Interference terms}
It is easy to see that if one looks on probability (frequency)
terms in different "measurement" classical games there are typical
quantum interference terms. For example for the game in
introduction one has in the first "measurement game"
$p_1=\cos^2\alpha,\; p_3=\sin^2\alpha$ but in the second
"measurement game" there are $p_2=\cos^2(\alpha-\theta_a),\;
p_4=\sin^2(\alpha-\theta_a) $ with fixed $\theta_a$ [12]. So
\[p_2=(\cos\alpha\cos\theta_a-\sin\alpha\sin\theta_a)^2=\cos^2\theta_a
p_1+ \sin^2\theta_a p_3-\sin2\theta_b \sqrt{p_1}\sqrt{p_3}.\] Same
can be seen for $p_4$.

This paper was partially supported by EU-Network
"QP and Applications''  and the Profile Mathematical
Modeling of V\"axj\"o University (A. A. Grib and A. Yu. Khrennikov)
 and Nat. Sc. Found., grant N PHY99-07949 at KITP, Santa-Barbara,
visiting professor fellowship
at Russian State Humanitarian  University (A. Yu. Khrennikov).

\medskip

\centerline{\bf REFERENCES}

\medskip

1. A. K. Ekert, {\it Phys. Rev. Lett.} {\bf  67}, 661 (1999).

2.  A. Yu. Khrennikov (editor), {\it Foundations of Probability and Physics,}
{\it Q. Prob. White Noise Anal.},  13,  WSP, Singapore, 2001.

3. A. Yu. Khrennikov (editor), {\it Quantum Theory: Reconsideration
of Foundations,} Ser. Math. Modeling, 2, V\"axj\"o Univ. Press,  2002.

4.  A. Yu. Khrennikov (editor), {\it Foundations of Probability and Physics}-2,
Ser. Math. Modeling, 5, V\"axj\"o Univ. Press,  2003.

5.  A. Yu. Khrennikov (editor),  {\it Quantum Theory: Reconsideration
of Foundations}-2,  Ser. Math. Modeling, 10, V\"axj\"o Univ. Press,  2004.

6.  A. Yu. Khrennikov, {\it J. Phys.A: Math. Gen.} {\bf 34}, 9965-9981 (2001).

7.   A. Yu. Khrennikov, {\it Information dynamics in cognitive, psychological and
anomalous phenomena,} Ser. Fundamental Theories of Physics, Kluwer, Dordreht, 2004.

8. A. Yu. Khrennikov,  {\it J. Math. Phys.} {\bf 44},  2471- 2478 (2003).

9.  G. Birkhoff, J. von Neumann,
{\it Ann. Math.} {\bf 37}, 823-843 (1936).

10. A. A. Grib, R. R. Zapatrin, {\it  Int. Journ. Theor. Phys.} {\bf  29}, 113-120 (1990).

11.  A. A. Grib, R. R. Zapatrin.
{\it Int. Journ. Theor. Phys.} {\bf  30},  949-955 (1991).

12. A. A. Grib, G. N. Parfionov, {\it Seminars POMI}
{\bf 291}, 1-24 (2002).

13. A. A. Grib, A. Yu. Khrennikov, K. A. Starkov, {\it ``Probability
amplitude in quantum like games''} in {\it Quantum
Theory: reconsideration of foundation} -- 2, edited by A. Yu. Khrennikov,
Ser. Math. Modeling, 10, V\"axj\"o Univ. Press,
2003, pp. 703-721.

14. A. A. Grib, Jr. W. A.Rodrigues, {\it Nonlocality in
Quantum Physics}, Kluwer Academic // Plenum Publishers, N.Y.,
Boston, Dodrecht, London, Moscow, 1999.

\end{document}